\documentclass[12pt]{article}
\usepackage{amssymb}
\usepackage{amsmath}

\numberwithin{equation}{section}

\newcommand{\ii}{\mathrm{i}}

\newcommand{\mz}{\mathbb{Z}}

\newcommand{\be}{\begin{equation}}
\newcommand{\ee}{\end{equation}}
\newcommand{\bea}{\begin{eqnarray}}
\newcommand{\eea}{\end{eqnarray}}
\newcommand{\ba}{\begin{array}}
\newcommand{\p}[1]{(\ref{#1})}
\newcommand{\ea}{\end{array}}

\def\bbox{{\,\lower0.9pt\vbox{\hrule \hbox{\vrule height 0.2 cm
        \hskip 0.2 cm \vrule height 0.2 cm}\hrule}\,}}
\newcommand{\dsl}{\pa \kern-0.5em /}

\newcommand{\cp}{{\mathbb{C}P}}

\def\CL{{\cal L}}                       

\begin{document}


\begin{titlepage}

\begin{flushright}
   Imperial/TP/03-04/9\\
   hep-th/0403038\\
\end{flushright}

\vskip 1cm

\begin{center}

\baselineskip=16pt

{\Large\bf A New Infinite Class\\[0.3cm]
  of Sasaki-Einstein Manifolds}

\vskip 1.3cm

Jerome P. Gauntlett$^{1*}$, Dario Martelli$^{2}$, James F.
   Sparks$^{2}$ and Daniel Waldram$^{2}$

\vskip 1cm

{\small{\it $^1$Perimeter Institute for Theoretical Physics\\
    Waterloo, ON, N2J 2W9, Canada\\
    E-mail: jgauntlett@perimeterinstitute.ca\\}} \vskip .6cm
{\small{\it$^{2}$Blackett Laboratory, Imperial College\\
    London, SW7 2BZ, U.K.\\
    E-mail: d.martelli, j.sparks, d.waldram@imperial.ac.uk \\}}

\vskip 2cm


\end{center}

\begin{center}
   \textbf{Abstract}
\end{center}

\begin{quote}

We show that for every positive curvature K\"ahler--Einstein
manifold in dimension $2n$ there is a countably infinite class
of associated Sasaki--Einstein manifolds $X_{2n+3}$ in dimension
$2n+3$. When $n=1$ we recover a recently discovered family of
supersymmetric $AdS_5\times X_5$ solutions of type IIB string theory,
while when $n=2$ we obtain new supersymmetric $AdS_4\times X_7$
solutions of $D=11$ supergravity. Both are expected to provide new
supergravity duals of superconformal field theories.

\end{quote}

\vfill \vskip 5mm
\hrule width 5.cm
\vskip 5mm
{\small{\noindent$^*$ On leave from: Blackett Laboratory, Imperial
  College, 
  London, SW7 2BZ, U.K.\\}}

\end{titlepage}

\section{Introduction}

A Sasaki--Einstein manifold $X_{2m+1}$ may be defined as a
complete Riemannian manifold of dimension $2m+1$ such that the
metric cone $C(X_{2m+1})$ is Ricci-flat and K\"ahler {\it i.e.}
Calabi--Yau. The restricted holonomy group of the cone is then a
subgroup of $SU(m+1)$. There are various other, more-or-less
equivalent, definitions. For example, $X_{2m+1}$ may also be
defined as a Riemannian manifold which admits two\footnote{When
$m=3$ one can have weak $G_2$ holonomy manifolds
  admitting a \emph{single} solution to the Killing spinor
  equation. In higher dimensions this is not possible -- see, {\it
    e.g.}~\cite{tri}.}
non-trivial solutions to the Killing spinor equation
$\nabla_X \psi = \pm\frac{1}{2}\ii X\cdot \psi$, with appropriate signs.
Alternatively, the Sasakian-Einstein condition may be phrased in
terms of the existence of a certain type of contact structure --
indeed, this was Sasaki's definition. These conditions are all
equivalent, provided $X_{2m+1}$ is simply-connected and spin
\cite{tri}.

There has been particular interest in Sasaki--Einstein manifolds
recently due to their importance in the AdS/CFT correspondence
\cite{Maldacena:1997re}. For example, $AdS_5\times X_5$, with
suitable self-dual five-form flux, is a supersymmetric solution of
type IIB supergravity that is expected to be dual to a
four-dimensional superconformal field theory arising from a stack of
D3-branes placed at the tip of the corresponding Calabi--Yau cone
\cite{Klebanov:1998hh,Morrison:1998cs}. The superconformal field
theories will preserve at least $\mathcal{N}=1$ supersymmetry
\cite{Klebanov:1998hh,Morrison:1998cs,Figueroa-O'Farrill:1998nb,
Acharya:1998db}. Similarly, $AdS_4\times X_7$, with appropriate
four-form flux, is a supersymmetric solution of eleven-dimensional
supergravity that is expected to be dual to a three-dimensional
superconformal field theory arising on a stack of M2-branes
sitting at the tip of the corresponding metric cone. The
superconformal field theories will preserve at least
$\mathcal{N}=2$ supersymmetry\footnote{If $X_7$ has weak $G_2$
holonomy, so that the metric cone has $Spin(7)$ holonomy, then the
$D=11$ solutions will be dual to field theories with the minimal
$\mathcal{N}=1$ supersymmetry, while $\mathcal{N}=3$
theories arise as duals to tri-Sasakian $X_7$.}
\cite{Klebanov:1998hh,Morrison:1998cs,Figueroa-O'Farrill:1998nb,
Acharya:1998db}.

Recall that every Sasaki--Einstein manifold possesses a
constant-norm Killing vector, $V$ (for some general discussion see
{\it e.g.} \cite{boyer}). Indeed, in the contact structure
definition, $V$ is part of the geometric data. We thus have a
one-dimensional foliation, and it is easy to show that the
transverse geometry is locally K\"ahler--Einstein of positive
curvature. If the orbits of $V$ close, then we have a $U(1)$
action. Since $V$ is nowhere vanishing, it follows that the
isotropy groups of this action are all finite. Thus the space of
leaves of the foliation will be a positive curvature
K\"ahler--Einstein orbifold of complex dimension $m$. Such
Sasaki--Einstein manifolds are called quasi-regular. If the $U(1)$
action is free, the space of leaves is actually a
K\"ahler--Einstein manifold and the Sasaki--Einstein manifold is
then said to be regular. Moreover, the converse is true: there is
a Sasaki--Einstein structure on the total space of a certain
$U(1)$ bundle over any given K\"ahler--Einstein manifold of
positive curvature \cite{Kob}. A similar result is true in the
quasi-regular case \cite{boyer}. If the orbits of $V$ do not
close, the Sasaki--Einstein manifold is said to be irregular.

Although there are many results in the literature on
Sasaki--Einstein manifolds in every dimension, explicit metrics
are rather rare. Homogeneous regular Sasaki--Einstein manifolds
are classified: they are all $U(1)$ bundles over generalised flag
manifolds \cite{tri}. This result follows from the classification
of homogeneous K\"ahler--Einstein manifolds. For example, in low
dimensions with $m=1,2,3$ the homogeneous base is either $\cp^1$,
$\cp^2$, $\cp^1\times \cp^1$, $\cp^3$, $\cp^2\times \cp^1$,
$\cp^1\times\cp^1\times\cp^1$, $SU(3)/T^2$ or $\mathrm{Gr}_{5,2}$.
The corresponding homogeneous Sasaki--Einstein manifolds are, in
the physics literature, denoted $S^3$, $S^5$, $T^{1,1}$, $S^7$,
$M^{3,2}$, $Q^{1,1,1}$, $N^{1,1}$, $V_{5,2}$, respectively (see
for instance \cite{Duff:hr} for a review). Notice that
inhomogeneous K\"ahler--Einstein manifolds are known to exist
(including cohomogeneity one examples), and are explicit up to a
function which satisfies a certain ODE -- see, {\it e.g.}
addendum C of \cite{Besse}. One may then construct the associated
regular Sasaki--Einstein manifolds.

However, until recently, there have been no known explicit
inhomogeneous metrics in the quasi-regular class. Previous
constructions\footnote{Very recently an infinite class of explicit
inhomogeneous Einstein metrics have been constructed in
\cite{Hashimoto:2004kc}. It is not known if any of these are
Sasaki--Einstein.} of such manifolds use methods in algebraic
geometry which imply existence of solutions, but do not produce
explicit metrics. Moreover, no irregular examples were known to
exist at all. In \cite{paper2}, we constructed a countably
infinite family of explicit cohomogeneity one Sasaki--Einstein
metrics on $S^2\times S^3$. 
These were discovered by dualising one of the
class of M-theory solutions of \cite{paper1}. This family gives
not only the first explicit examples of inhomogeneous
quasi-regular Sasaki--Einstein manifolds, but also the first
examples of irregular geometries. In this note we show that this
construction extends to every dimension. More precisely, we prove
the following:

\bigskip

\noindent \textsl{For every positive curvature $2n$-dimensional
K\"ahler--Einstein manifold $B_{2n}$, there is a countably
infinite class of associated compact, simply-connected, spin,
Sasaki--Einstein manifolds $X_{2n+3}$ in dimension $2n+3$. Hence
each $X_{2n+3}$ admits a Killing spinor and the holonomy of the
associated metric cone is a subgroup of $SU(n+2)$. Moreover, the
isometry group of $X_{2n+3}$ is at least $G\times T^2$, where $G$
is the isometry group of $B_{2n}$.}

\bigskip 

\noindent Note that there is also a straightforward generalisation
to the case where $B_{2n}$ is a product of K\"ahler--Einstein
manifolds~\cite{stras}.   

\section{The local metric}

Let $B_{2n}$ be a (complete) $2n$-dimensional positive curvature
K\"ahler--Einstein manifold, with metric $d\tilde s^2$ and K\"ahler
form $\tilde J$. Recall that complete Einstein manifolds with
positive curvature are necessarily compact \cite{Myers} and hence
$B_{2n}$ is compact. In addition, we note that a positive
curvature K\"ahler--Einstein manifold is simply-connected
\cite{Kob}. Now, as shown in \cite{BB}, \cite{PP}, the local
metric
\bea d\hat s^2 & = & U(\rho)^{-1} d\rho^2 + \rho^2 U(\rho)
(d \tau - A )^2 +\rho^2 d\tilde s^2 \label{metric} \eea is then a
positive curvature K\"ahler--Einstein metric in dimension $2n+2$,
where $2\tilde J = dA$ and
\bea
   U(\rho) = \frac{\lambda}{2n+2}
      - \frac{\Lambda}{2n+4}\rho^2
      + \frac{\Lambda}{2(n+1)(n+2)} \left(\frac{\lambda}
         {\Lambda}\right)^{n+2}\frac{\kappa}{\rho^{2n+2}}
\eea
where
$\kappa$ is a constant that has been rescaled compared to
\cite{PP} for later convenience. The K\"ahler form is given by
\bea \hat J & = & \rho^2 \tilde J +  \rho (d\tau - A) \wedge d\rho
\eea
and the curvatures of the K\"ahler--Einstein metrics in dimensions
$2n$ and $2n+2$ are $\lambda>0$ and $\Lambda>0$, respectively.
Thus $\widetilde{\mathrm{Ric}} = \lambda \tilde{g}$,
$\widehat{\mathrm{Ric}} = \Lambda \hat{g}$.

In \cite{PP} it was shown that the local expression (\ref{metric})
describes a complete metric on a manifold if and only if the base
is $\cp^n$, with its canonical metric, and the resulting total
space is then $\cp^{n+1}$, again with canonical metric. This
requires the choice $\kappa=0$ \cite{PP}. For this reason, the
local one-parameter family of K\"ahler--Einstein metrics
(\ref{metric}) (with $\Lambda>0$) has been largely ignored for
more than 20 years.

However, consider adding another dimension to the metric above --
specifically, the local Sasaki--Einstein direction. We define the
$(2n+3)$-dimensional local metric
\bea
   ds^2 = d\hat s^2 + (d\psi^{\prime} + \sigma)^2
\label{ES}
\eea
where $d\sigma = 2\hat J$. As is well-known (see,
{\it e.g.} \cite{hartnoll} for a recent review), such a metric is
locally Sasaki--Einstein, with curvature $2n+2$, provided
$\Lambda=2(n+2)$. An appropriate choice for the connection
one-form $\sigma$ is
\bea \sigma = \frac{\lambda}{\Lambda}A +
   \left(\frac{\lambda}{\Lambda}-\rho^2\right) (d\tau - A)~.
\eea

We now utilise a very useful change of coordinates which casts the
local metric (\ref{ES}) into a different $(2n+2)+1$ decomposition.
Define the new coordinates
\bea
   \alpha = - \tau - \frac{\Lambda}{\lambda}\psi^{\prime}~
\eea
and $(\Lambda/\lambda)\psi^{\prime} = \psi$. A straightforward
calculation then shows that the local metric (\ref{ES}) assumes
the following form
\bea
   ds^2 = U(\rho)^{-1}d\rho^2 + \rho^2d\tilde s^2
      + q(\rho) (d\psi + A)^2
      + w(\rho) \left[d\alpha + f(\rho)(d\psi +A)\right]^2
\label{new} \eea where\footnote{Note that for $n=1$ these definitions
   differ slightly from those in \cite{paper2} when using the
   coordinates in section 5.} 
\bea
   w(\rho) & = & \rho^2 U(\rho)+(\rho^2-\lambda/\Lambda)^2\\
   q(\rho) & = &
      \frac{\lambda^2}{\Lambda^2}\frac{\rho^2U(\rho)}{w(\rho)}\\
   f(\rho) & = &
      \frac{\rho^2(U(\rho)+\rho^2-\lambda/\Lambda)}{w(\rho)}~.
\label{deff}
\eea
In order to have a Riemannian metric we must certainly ensure that
$U\geq0$, $w \geq 0$, $q \geq 0$. The second two conditions are in
fact implied by the first, $U\geq 0$. The latter holds if we
choose the range of $\rho$ to be
\be
   \rho_1 \leq \rho \leq \rho_2
\ee
where $\rho_i$ are two appropriate roots of the equation
$U(\rho)=0$. As we want to exclude $\rho=0$, since the metric is
generically singular there, we thus take $\rho_i$ to be both positive
(without loss of generality).
Moreover, $w>0$ provided $\rho_i^2\neq \lambda/\Lambda$. If we
define
\be x=\frac{\Lambda}{\lambda}\rho^2 \ee
we can write
\be
U(\rho)=\frac{\lambda}{2(n+1)(n+2)}\frac{1}{x^{n+1}}P(x;\kappa)
\ee
where we have introduced the polynomial in $x$
\bea\label{polyx}
P(x;\kappa) = -(n+1)x^{n+2}+(n+2)x^{n+1}+\kappa~.
\eea
For any $x\geq 0$ (for fixed $\kappa$), a zero of $U(\rho)$
corresponds to a root of the equation $P(x;\kappa)=0$. Observe
that the only turning points of $P(x;\kappa)$ on the interval
$[0,\infty)$ are at $x=0$ and $x=1$ where the latter is a global
maximum. When $\kappa$ is positive it is straightforward to see
that there is only a single real root and so this is excluded.
Similarly, when $\kappa<-1$ there are no real roots and this is
also excluded. We thus conclude that in the range
\bea -1<\kappa\le 0 \eea
there are two distinct suitable values of the roots $\rho_i$
\bea
    0 \leq \rho_1 < \sqrt{\tfrac{\lambda}{\Lambda}}
       < \rho_2 \leq \sqrt{\tfrac{\lambda (n+2)}{\Lambda(n+1)}}~.
\eea
The limiting value $\kappa=0$ in \p{metric} gives rise to a
smooth compact K\"ahler--Einstein manifold if and only if
$B_{2n}=\cp^n$, in which case the resulting space is $\cp^{n+1}$.
Clearly the corresponding Sasaki--Einstein manifold $X_{2n+3}$, defined
by \p{ES}, is
$S^{2n+3}$ (or a discrete quotient thereof). As a consequence, we now
focus on the case where the range of $\kappa$ is
$-1<\kappa<0$.

\section{Global analysis}

We first analyse the $(2n+2)$-dimensional part of (\ref{new}),
transverse to the $\alpha$ direction, and show that, with
appropriate ranges and identifications of the coordinates, it is a
smooth complete metric on the total space of an $S^2$ bundle over
the original $2n$-dimensional K\"ahler--Einstein manifold $B_{2n}$.
This is true for any value of the constant $\kappa$, with $-1
<\kappa<0$. We will also show that the resulting
$(2n+2)$-dimensional Riemannian manifold $Y_{2n+2}$ is in fact
conformally K\"ahler. Secondly, we show that the $\alpha$
direction can be made into a $U(1)$ fibration over the base
$Y_{2n+2}$ for countably infinitely many values of the parameter
$\kappa$ in the allowed range. The total space of the fibration is
then a $(2n+3)$-dimensional Einstein manifold. In the following
section we show that it is Sasaki--Einstein.

\subsection*{The base}

With the range of $\rho$ being $\rho_1 \leq \rho \leq \rho_2$ we
see that, if $\psi$ is periodically identified, the $\rho-\psi$
part of the metric (\ref{new}) at fixed point on the base $B_{2n}$
is a circle fibred over the line segment with coordinate $\rho$,
where the size of the circle goes to zero at the two roots
$\rho=\rho_i$. This fibre will be a smooth $S^2$ provided the
period of $\psi$ can be chosen so that there are no conical
singularities where the circle collapses. Now, near a root
$\rho_i$ we have $U(\rho)\approx U'(\rho_i) (\rho-\rho_i)$.
Defining the new coordinate
\bea
R^2 = \frac{4(\rho-\rho_i)}{U'(\rho_i)}
\eea
the $\rho-\psi$ part of the metric (\ref{new}), at any fixed point on
the base, is
\bea
dR^2 + \frac{\lambda^2}{4\Lambda^2}
   \frac{\rho_i^2U'(\rho_i)^2}{(\rho_i^2-\lambda/\Lambda)^2} R^2
   d\psi^2
\eea
near to the root $\rho=\rho_i$. Now we notice the
remarkable fact that
\bea
\frac{\rho_i^2U'(\rho_i)^2}{(\rho_i^2-\lambda/\Lambda)^2} =
   \Lambda^2~
\eea
for any root. Therefore the circle fibre collapses smoothly at both
roots provided that $\psi$ has period $\frac{4\pi}{\lambda}$. The
$\rho-\psi$ fibre is then topologically $S^2$.

Now, since $dA=2\tilde{J}$ and $\widetilde{\mathrm{Ric}} = \lambda
\tilde{g}$, we see that $dA$ is $\frac{2}{\lambda}$ times the Ricci form
of the base manifold $B_{2n}$. Since the period of $\psi$ is
$\frac{4\pi}{\lambda}$, at fixed value of $\rho$ between the
roots, the resulting circle fibration is just the associated
$U(1)$ bundle to the canonical bundle of $B_{2n}$. Indeed, note
that one may rescale the coordinate $\rho$, as well as the other
variables and coordinates, so as to set $\lambda=2$. Then the
period of $\psi$ is the canonical value of $2\pi$, and $dA$ is the
Ricci form. We may therefore set
\begin{equation}
\begin{aligned}
   \lambda &= 2 \\
   \Lambda &= 2(n+2)
\end{aligned}
\end{equation}
without loss of generality, where the second
condition ensures that the Sasaki--Einstein metric has curvature
$2(n+1)$. We will often still keep factors of $\lambda$ and
$\Lambda$ in formulae, with the understanding that they take the
above values.

Before studying further topological properties of these spaces, we
remark that the spaces $Y_{2n+2}$ are complex manifolds. Moreover,
these manifolds are in fact conformally K\"ahler. Specifically,
consider the fundamental two-forms
\bea
J_{\pm}  =  \rho^2 \tilde J
   \pm \frac{\lambda}{\Lambda}\frac{\rho}{\sqrt{w(\rho)}}\;
   d\rho\wedge (d\psi +A )
\eea
and the associated $(n+1,0)$-forms
\bea
\Omega_{\pm} = \rho^n \tilde\Omega \wedge \left[
   \frac{d\rho}{\sqrt{U(\rho)}}
   \pm i \sqrt{q(\rho)} \left(d\psi +A\right) \right]~.
\eea
One finds that one can write $d\Omega^\pm$ as
\bea
d\Omega_\pm = L_\pm \wedge\Omega_\pm
\eea
for suitable one-forms $L_\pm$, which shows that
$Y_{2n+2}$ is complex, with complex structure specified by
$\Omega_{\pm}$. Moreover, it is immediate that
\bea
dJ_\pm = \frac{2}{\rho}\left(
   1 \mp \frac{\lambda}{\Lambda}\frac{1}{\sqrt{w(\rho)}}
   \right) d\rho\wedge J_\pm
\eea
which implies that $Y_{2n+2}$ is conformal to a K\"ahler manifold.

We have now constructed a complete manifold $Y_{2n+2}$ which is
the total space of an $S^2$ bundle over $B_{2n}$. Let us denote
the canonical bundle of $B_{2n}$ as $\CL$. Then $Y_{2n+2}$ may be
thought of as the manifold obtained by adding a point to each
$\mathbb{C}$ fibre of $\CL$ to obtain a $\cp^1=S^2$ bundle over
$B_{2n}$. We can write this bundle as $\CL \times_{U(1)} \cp^1$. This
notation means that one uses the $U(1)$ 
transition functions of $\CL$ to make an associated $\cp^1$
bundle, where $U(1)$ acts isometrically on $\cp^1$ in the usual
way. Notice that one can also write this as
$\mathbb{P}(\CL\oplus\mathcal{O})$, where $\mathcal{O}$ is a
trivial complex line bundle over $B_{2n}$ and $\mathbb{P}$ denotes
projectivisation ({\it i.e.} one quotients out by the Hopf map
$\mathbb{C}^2\mapsto \mathbb{C}^2/\mathbb{C}_*=\cp^1=S^2$ on each
$\mathbb{C}^2$ fibre of the bundle $\CL\oplus \mathcal{O}$).

We will later need a basis for the second homology group of
$Y_{2n+2}$, in terms of a basis for the base manifold $B_{2n}$. In
general we have $H_2(B_{2n};\mz)\cong
\mz^r\oplus\mathrm{torsion}$, where $r$ is the rank. We will not
need to know anything about the finite part (as we will be
integrating two-forms over the basis). Since $\pi_1(B_{2n})$ is
trivial, note that, by the universal coefficients theorem, we have
$H^2(B_{2n};\mz)\cong \mz^r$ is torsion-free. Now, we can find $r$
submanifolds $\Sigma_i$, $i=1,\ldots,r$, such that their homology
classes $[\Sigma_i]$ provide a basis for the free part of the
second homology group of $B_{2n}$. Recall that $Y_{2n+2}$ is the
total space of a $\cp^1=S^2$ bundle over $B_{2n}$. In fact, since
this is a projectivised bundle, we can use the results of section
20 of \cite{bott} to write down the cohomology ring of $Y_{2n+2}$
in terms of $B_{2n}$. In particular, we note that
$H^2(Y_{2n+2};\mz)\cong\mz\oplus H^2(B_{2n};\mz)$, where the first
factor is generated by the cohomology class of the $S^2$ fibre. By
Poincar\'e duality, notice that we can find $r$ cohomology classes
$\omega_i\in H^2(B_{2n};\mz)$, $i=1,\dots,r$, such that
$\left<\omega_i,[\Sigma_j]\right>=\delta_{ij}$. Also notice that
the rank of $H_2(Y_{2n+2};\mz)$ is then $r+1$. Now let
$\sigma^N:B_{2n}\rightarrow Y_{2n+2}$ denote the section of
$\pi:Y_{2n+2}\rightarrow B_{2n}$ corresponding to the ``north
pole" of the $S^2$ fibres. This is clearly a global section. Also
define a submanifold $\Sigma\cong S^2$ of $Y_{2n+2}$ corresponding
to the fibre of $Y_{2n+2}$ at some fixed point on the base
$B_{2n}$. Then since
$\left<\pi^*\omega_i,\sigma^N_*[\Sigma_j]\right>
=\left<\omega_i,[\Sigma_j]\right>=\delta_{ij}$, we see that
$\{\Sigma,\sigma^N\Sigma_i\}$ forms a representative basis for the
free part of $H_2(Y_{2n+2};\mz)$.

Now, since $Y_{2n+2}$ is simply-connected, note that we have
\begin{equation}\begin{aligned}
H^2(Y_{2n+2};\mathbb{Z})
      & \cong \left\{
         [\nu]\in H^2_{\mathrm{de\ Rham}}(Y_{2n+2}):
         \int_Z\nu \in \mathbb{Z},
         \ \forall Z\subset Y_{2n+2}, \ \partial Z=\emptyset
      \right\} \\ & \equiv H^2_{\mathrm{de\
      Rham}}(Y_{2n+2};\mz)~.\end{aligned}\end{equation}
Thus a closed two-form $\nu$ corresponds to an integral cohomology
class under the map $H^2(Y_{2n+2};\mz)\rightarrow
H^2(Y_{2n+2};\mathbb{R})$ if and only if the periods
$\left<\nu,[Z]\right>$ over our basis
$Z\in\{\Sigma,\sigma^N\Sigma_i\}$ are all integral.

We conclude this section with two technical comments. First, note
that $Y_{2n+2}$ is always a spin manifold, irrespective of whether
$B_{2n}$ is spin or not. One way to see this is to view $Y_{2n+2}$
as the total space of the unit sphere bundle in the associated
$\mathbb{R}^3$ bundle, $E$. Then
$w_2(E)=w_2(\CL_{\mathbb{R}})=c_1(\CL) \mod 2$. But also
$w_2(B_{2n})=c_1(B_{2n}) \mod 2 =-c_1(\CL) \mod 2$. We may then use the
Whitney formula to compute $w_2(TE)\mid_{B_{2n}} =
w_2(E)+w_2(B_{2n}) = 0$, where we view $E$ as the normal bundle to
the zero section $B_{2n}$ of $E$. Since $B_{2n}$ is a deformation
retract of the total space of $E$, this is enough to show that the
total space of $E$, and therefore its boundary $Y_{2n+2}$, are
spin. Secondly, the $S^2$ bundle $Y_{2n+2}$ will not in general be
trivial. It was shown in \cite{paper2} that if $n=1$ and $B_2=S^2$
then $Y_4\cong S^2\times S^2$, but this is a low-dimensional
exception. Using standard techniques one can show that the first
Pontryagin class of the $SO(3)$ bundle $Y_{2n+2}$ is
$c_1^2(\CL)\in H^4(B_{2n};\mz)$. In fact this can never be trivial
for $n\geq2$ since, for a K\"ahler--Einstein manifold $B_{2n}$,
$c_1(\CL)$ is proportional to the K\"ahler class $[\tilde{J}]$,
and $\tilde{J}^n$ is proportional to the volume form, which is a
non-trivial cohomology class. In particular, $c_1^2(\CL)$ must be
non-trivial.

\subsection*{The circle fibration}

We now turn to the $\alpha$ direction in the metric (\ref{new}).
The latter may be written
\bea ds^2 = ds^2_{Y_{2n+2}}+w(\rho)(d\alpha+B)^2\eea
where $B=f(\rho)(d\psi+A)$. Since for the range of $\kappa$ chosen
the function $w>0$, the Killing vector $\partial/\partial\alpha$
is nowhere vanishing on the base $Y_{2n+2}$. The idea now is that,
if $\ell^{-1}B$ is a connection one-form, where $\ell\in \mathbb{R}$ is
some real number, then by periodically identifying $\alpha$ with
period $2\pi \ell$ we obtain a complete metric on the total space of
the $U(1)$ bundle over $Y_{2n+2}$ with connection one-form
$\ell^{-1}B$. We will find that a countably infinite number of values
of $\kappa$ work.

Notice that at fixed value of $\rho$ between the two roots, $B$ is
proportional to the global angular form $\frac{1}{2\pi}(d\psi+A)$ on the
resulting ``equatorial" $U(1)$ bundle. This is of course globally
defined on this space. However, crucially, the function $f(\rho)$
does not vanish at the poles. Since the azimuthal angle $\psi$ is
not defined at the poles the one-form $B$ is not well-defined on
the whole manifold $Y_{2n+2}$. However, it is straightforward to
verify that $dB$ is a globally defined smooth two-form on
$Y_{2n+2}$. We now address the question of when such a closed
two-form is the curvature of a connection on a complex line
bundle. In fact, this is the case precisely if the periods of the
two-form are integral. Indeed, we have
\bea
   \mathrm{Vect}_1^{\mathbb{C}}(Y_{2n+2})
      \cong H^2(Y_{2n+2};\mathbb{Z})
      \cong H^2_{\mathrm{de\ Rham}}(Y_{2n+2};\mz)
\eea
where the notation $\mathrm{Vect}_1^{\mathbb{C}}(Y_{2n+2})$
denotes isomorphism classes of complex line bundles over
$Y_{2n+2}$, and the second isomorphism is true only for a
simply-connected manifold $Y_{2n+2}$ (otherwise one misses the
torsion). Any closed two-form $\nu$, with $[\nu]\in
H^2_{\mathrm{de\ Rham}}(Y_{2n+2};\mz)$, is then the curvature of a
$U(1)$ bundle (divided by $2\pi$). Thus we must ensure that the
periods of $\frac{1}{2\pi}\ell^{-1}dB$ are integers, for some $\ell\in
\mathbb{R}$. In this case, $\ell^{-1}B$ is a local connection
one-form. The reason that it is not defined globally
(specifically, at the poles) is that, for a non-trivial bundle,
the connection $\ell^{-1}B$ is only defined in local coordinate
patches, and there are gauge transformations at the intersections
of the patches. The singular nature of $B$ is due to the fact that
we are trying to use only one patch for a non-trivial bundle.

The periods of $\frac{1}{2\pi}dB$ over our representative basis
$\{\Sigma,\sigma^N\Sigma_i\}$ for the free part of
$H_2(Y_{2n+2};\mz)$ are easy to compute:
\bea \int_{\Sigma} \frac{dB}{2\pi} = f(\rho_1)-f(\rho_2)\eea
\bea \int_{\sigma^N\Sigma_i} \frac{dB}{2\pi} = f(\rho_2)
c_{(i)}~.\eea
Here
\bea
c_{(i)}=\int_{\Sigma_i}\frac{dA}{2\pi}
   =\left<c_1(\CL),[\Sigma_i]\right>\in\mz
\eea
are the Chern numbers. Thus we see that, if
$f(\rho_1)/f(\rho_2)=p/q\in\mathbb{Q}$ is rational (with
$p,q\in\mathbb{Z}$), then the periods of $\left[q/2\pi
f(\rho_2)\right]dB$ are $\{p-q, qc_{(i)}\}$. Clearly, these are
all integer. Moreover, if we set $h=\mathrm{hcf}\{p-q, qc_{(i)}\}$
then $\frac{1}{2\pi}\ell^{-1}dB$ has integral periods with no common
factor, where $\ell=hf(\rho_2)/q$. Notice that, using the expression
(\ref{deff}), we have
\bea
R(\kappa)\equiv\frac{f(\rho_1)}{f(\rho_2)}  =
\frac{\rho^2_1(\rho_2^2 - \lambda/\Lambda)}
{\rho^2_2(\rho^2_1-\lambda/\Lambda)}=\frac{x_1(x_2-1)}{x_2(x_1-1)}~,
\eea
where $x_i$ are the associated roots of the polynomial
\p{polyx}. This is a continuous function of $\kappa$ in the
interval $(-1,0]$. Moreover, it is easy to see that $R(0)=0$ and
in the limit $R(-1)=-1$, and hence there are clearly a countably
infinite number of values of $\kappa$ for which $R(\kappa)$ is
rational and equal to $p/q$, with $|p/q|< 1$, and these all give
complete Riemannian manifolds.

\section{The global Sasaki structure on $X_{2n+3}$}

We denote the manifold with metric (\ref{new}), and appropriate
value of $\kappa$, by $X_{2n+3}$. By construction, this is a
complete, compact Einstein manifold. Moreover, since $X_{2n+3}$ is
the total space of a $U(1)$ fibration over a simply-connected
manifold $Y_{2n+2}$, it is straightforward to calculate
$\pi_1(X_{2n+3})$. Indeed, notice that, again by construction, the
Chern numbers have no common factor, and hence it follows that
$X_{2n+3}$ is also simply-connected\footnote{Look at the
Thom-Gysin sequence for the $U(1)$ fibration -- {\it cf}. appendix
A of \cite{paper2}. }. Note also that $w_2(X_{2n+3})$ is just the
pull-back of $w_2(Y_{2n+2})=0$, and so $X_{2n+3}$ is a spin
manifold.

The local metric (\ref{ES}) that we started with is
Sasaki--Einstein. The Sasaki condition means that there exists a
certain contact structure. This may be phrased in terms of the
existence of certain geometric objects which satisfy various
algebraic and differential conditions. Specifically, the contact
structure involves the Killing vector
$V=\partial/\partial\psi^{\prime}$, its dual $\chi$, and the
$(1,1)$ tensor constructed by raising an index of $d\chi$ with the
metric. Since these objects already locally satisfy the
appropriate algebraic and geometric conditions, we simply have to
check that they are globally well-defined, given the
identifications and ranges we have taken in order to make a
complete manifold. 

In terms of the coordinates (\ref{new}) the unit-norm vector $V$ is
easily computed to be 
\bea V =
   \frac{\Lambda}{\lambda}\left(\frac{\partial}{\partial\psi}
   -\frac{\partial}{\partial\alpha}\right)~.
\eea
This is clearly globally defined, as each of the vectors
$\partial/\partial\psi$, $\partial/\partial\alpha$ is globally
defined. Moreover, the dual one-form $\chi$ may be written
\bea \chi = \left(\rho^2-\frac{\lambda}{\Lambda}\right)[d\alpha +
f(d\psi+A)]+\frac{\Lambda}{\lambda}q(\rho)(d\psi+A)~.\eea
The function $(\rho^2-\lambda/\Lambda)$ is smooth on $X_{2n+3}$,
and the one-form $[d\alpha+f(d\psi+A)]$ is proportional to the
global angular form on the $U(1)$ fibration $U(1)\hookrightarrow
X_{2n+3}\rightarrow Y_{2n+2}$, so is globally defined on
$X_{2n+3}$. The global angular form $\frac{1}{2\pi}(d\psi+A)$ on the
``equatorial $U(1)$ bundle" contained in $Y_{2n+2}$ at fixed 
$\rho$, $\rho_1<\rho<\rho_2$, is not defined at the poles of the
$S^2$ fibres of $Y_{2n+2}$, but the function $q(\rho)$ vanishes
smoothly there, so in fact the second term is also globally
defined on $X_{2n+3}$. Clearly $\chi$ and also $d\chi$ are smooth
and globally defined. We conclude that the local Sasaki structure
is globally well-defined, and thus $X_{2n+3}$ is a
simply-connected spin Sasaki--Einstein manifold.

As discussed in \cite{boyer}, the restricted holonomy group of the
metric cone on a Sasaki--Einstein manifold $X_{2n+3}$ is a subgroup
of $SU(n+2)$. Since our manifolds $X_{2n+3}$ are simply-connected,
it follows that the metric cone $C(X_{2n+3})$ has holonomy group
contained in $SU(n+2)$, and is therefore a Calabi--Yau cone. Thus, from
\cite{bar}, there exists a parallel spinor on
$C(X_{2n+3})$ which projects down to a Killing spinor on
$X_{2n+3}$.

Notice that the orbits of the Killing vector $V$ close if and only
if $f(\rho_2)\in\mathbb{Q}$. In this case the Sasaki--Einstein
manifold is said to be quasi-regular, which means that the space
of leaves of the foliation determined by $V$ is a
K\"ahler--Einstein orbifold. Determining when
$f(\rho_2)\in\mathbb{Q}$ seems to be a non-trivial
number-theoretic problem. For the case $n=1$ studied in
\cite{paper2} one has to solve a quadratic diophantine, which can
be done using standard methods. However, there is no general
approach for solving higher-order diophantines. If $f(\rho_2)$ is
irrational, the orbits of $V$ are dense in the torus defined by
the Killing vectors $\partial/\partial\psi$,
$\partial/\partial\alpha$. In this case the Sasaki--Einstein
manifold is said to be irregular, of rank 2. There is no
well-defined K\"ahler--Einstein base in this case. These remarks
are of course consistent with \cite{PP} where it was shown that
the local K\"ahler--Einstein metric (\ref{metric}) is a complete
metric on a manifold only if $\kappa=0$. For the countably
infinite number of values of $\kappa$ found here, the
K\"ahler--Einstein ``base" is at best an orbifold, and in the
irregular case there is in fact no well-defined base at all.

\section{Connection with previous results}

If we introduce the new coordinates
$\rho^2=(\lambda/\Lambda)(1-cy)$ and $\beta=c\tau$, the  metric
(\ref{metric}) can be cast in the following form \be
d\hat{s}^2=\frac{\lambda}{\Lambda}(1-cy)d\tilde
s^2+\frac{1}{F(y)}dy^2
+\left(\frac{\lambda}{2\Lambda}\right)^2F(y)(d\beta-c A)^2 \ee
where \be
F(y)=\frac{2\Lambda}{(n+1)(n+2)}\frac{[a-y^2Q_n(1-cy)]}{(1-cy)^n}
\ee with the constant $a\equiv (\kappa+1)/c^2$ and $Q_n$ is a
polynomial of degree $n$ given by \bea Q_n(x)=\sum_{i=0}^n
(i+1)x^i~.\eea The range of $a$, for $c=1$, is given by $0<a\le
1$.

We are now in a position to recover the form of the
five-dimensional metrics, as presented in \cite{paper2}. Indeed,
this is achieved by setting $n=1$, $\lambda=2$ and $\Lambda=6$ to
obtain \be F(y)=\frac{2(a-3y^2+2cy^3)}{1-cy}~. \ee As a further
example, we note that seven-dimensional Sasaki--Einstein metrics
corresponding to $n=2$, $\lambda=2$ and $\Lambda=8$ have
\be
F(y)=\frac{\frac{4}{3}a-8y^2+\frac{32}{3}cy^3-4c^2y^4}{(1-cy)^2}
\ee

When $c$ is non-zero we can always choose $c=1$ by rescaling $y$.
On the other hand when $c=0$ we can recover known Sasaki--Einstein
metrics. Setting $c=0$, using the fact that
$Q_n(1)=\frac{1}{2}(n+2)(n+1)$,
and introducing the new coordinates $y^2=(2a/\Lambda)\cos^2\omega$
and $\beta'=[a\lambda^2/2(n+2)(n+1)]^{1/2}\beta$, we find that the
metric (\ref{metric}) becomes
\be
d\hat{s}^2=\frac{\lambda}{\Lambda}d\tilde s^2
   +\frac{1}{\Lambda}\left[d\omega^2+\sin^2\omega (d\beta')^2\right]
\ee
which is just the direct product metric on $B_{2n}\times S^2$. The
corresponding Sasaki--Einstein metrics (\ref{ES}) are then
appropriate $U(1)$ bundles over $B_{2n}\times S^2$ and are clearly
regular. For instance, with $n=1$ so that $B_2=\cp^1$, the $c=0$
limit gives the homogeneous space $T^{1,1}/\mz_2$ (or
$T^{1,1}$)~\cite{paper2}. Similarly, for $n=2$, explicit metrics are
obtained by taking $B_4$ to be $\cp^1\times \cp^1$ or $\cp^2$, and the
corresponding homogeneous Sasaki--Einstein seven-manifolds are
$Q^{1,1,1}$ and $M^{3,2}$, respectively. The new Sasaki--Einstein
metrics in seven dimensions that we have presented here, for the
special cases  with $B_4$ given by  $\cp^1\times \cp^1$ and
$\cp^2$, are co-homogeneity one generalisations of $Q^{1,1,1}$ and
$M^{3,2}$, respectively.

\subsection*{Acknowledgements}
DM is funded by an EC Marie Curie Individual Fellowship under contract number
HPMF-CT-2002-01539. JFS is funded by an EPSRC mathematics
fellowship. DW is supported by the Royal Society through a
University Research Fellowship.


\begin{thebibliography}{99}


\bibitem{tri} C.P. Boyer, K. Galicki, ``3-Sasakian Manifolds", Surveys
   Diff. Geom.{\bf 7} (1999) 123-184, [arXiv:hep-th/9810250].

\bibitem{Maldacena:1997re}
J.~M.~Maldacena, ``The large N limit of superconformal field
theories and supergravity", Adv.\ Theor.\ Math.\ Phys.\  {\bf 2},
231 (1998) [Int.\ J.\ Theor.\ Phys.\  {\bf 38}, 1113 (1999)]
[arXiv:hep-th/9711200].

\bibitem{Klebanov:1998hh}
I.~R.~Klebanov and E.~Witten, ``Superconformal field theory on
threebranes at a Calabi--Yau  singularity", Nucl.\ Phys.\ B {\bf
536}, 199 (1998) [arXiv:hep-th/9807080].

\bibitem{Morrison:1998cs}
D.~R.~Morrison and M.~R.~Plesser, ``Non-spherical horizons I",
Adv.\ Theor.\ Math.\ Phys.\  {\bf 3} (1999) 1
[arXiv:hep-th/9810201].

\bibitem{Figueroa-O'Farrill:1998nb}
J.~M.~Figueroa-O'Farrill, ``Near-horizon geometries of
supersymmetric branes", [arXiv:hep-th/9807149].

\bibitem{Acharya:1998db}
B.~S.~Acharya, J.~M.~Figueroa-O'Farrill, C.~M.~Hull and B.~Spence,
``Branes at conical singularities and holography", Adv.\ Theor.\
Math.\ Phys.\  {\bf 2} (1999) 1249 [arXiv:hep-th/9808014].

\bibitem{boyer}
C.~P.~Boyer and K.~Galicki,
``On Sasakian-Einstein Geometry'',
Internat. J. Math.  {\bf 11}  (2000), no. 7, 873--909,
[arXiv:math.DG/9811098].

\bibitem{Kob} S.~Kobayashi, ``Topology of positively pinched
K\"ahler manifolds", T\^ohoku Math. J. {\bf 15} (1963) 121-139.

\bibitem{Duff:hr}
M.~J.~Duff, B.~E.~W.~Nilsson and C.~N.~Pope, ``Kaluza--Klein
Supergravity", Phys.\ Rept.\  {\bf 130}, 1 (1986).

\bibitem{Besse} A.L. Besse, ``Einstein Manifolds",
Springer-Verlag, 2nd edition, 1987.

\bibitem{Hashimoto:2004kc}
Y.~Hashimoto, M.~Sakaguchi and Y.~Yasui,
``New infinite series of Einstein metrics on sphere bundles from AdS black
holes,''
arXiv:hep-th/0402199.

\bibitem{paper2}
J.~P.~Gauntlett, D.~Martelli, J.~F.~Sparks and D.~Waldram,
``Sasaki--Einstein metrics on $S^2\times S^3$",
[arXiv:hep-th/0403002].

\bibitem{paper1}
J.~P.~Gauntlett, D.~Martelli, J.~F.~Sparks and D.~Waldram,
``Supersymmetric $AdS_5$ solutions of M-Theory",
[arXiv:hep-th/0402153].

\bibitem{stras}
J.~P.~Gauntlett, D.~Martelli, J.~F.~Sparks and D.~Waldram,
``New $AdS$ Solutions in String and M-Theory", to be published in
\textit{Proceedings of the 73rd Meeting between Theoretical Physicists and
Mathematicians: the (A)dS-CFT correspondence}, 11--13 September, 2003,
Strasbourg, France. 

\bibitem{Myers}
S.B. Myers, Duke Math J. {\bf 8} (1941) 401.

\bibitem{BB}
L.~B\'erard-Bergery, Institut Elie Cartan {\bf 6}, 1 (1982).

\bibitem{PP}
D.~N.~Page and C.~N.~Pope, ``Inhomogeneous Einstein Metrics On
Complex Line Bundles", Class.\ Quant.\ Grav.\  {\bf 4}, 213
(1987).

\bibitem{hartnoll}
G.~W.~Gibbons, S.~A.~Hartnoll and C.~N.~Pope,
 ``Bohm and Einstein--Sasaki metrics, black holes and cosmological event
horizons", Phys.\ Rev.\ D {\bf 67} (2003) 084024
[arXiv:hep-th/0208031].

\bibitem{bott} R. Bott and L. Tu, ``Differential Forms in Algebraic
Topology", Springer-Verlag 1982.

\bibitem{bar} C. B\"ar, ``Real Killing spinors and holonomy", Comm.
Math. Phys. {\bf 154} (1993) 509-521.

\end{thebibliography}
\end{document}